# Optical property modification of ZnO: Effect of 1.2 MeV Ar irradiation

Soubhik Chattopadhyay[1], Sreetama Dutta[1], Palash Pandit[2], D. Jana[*,1], S. Chattopadhyay[3], A. Sarkar[4], P. Kumar[5], D. Kanjilal[5], D. K. Mishra[6], S. K. Ray[7]

[1]Department of Physics, University of Calcutta, 92 Acharya Prafulla Chandra Road, Kolkata-700 009, India.
[2]Department of Chemistry, University of Calcutta, 92 Acharya Prafulla Chandra Road, Kolkata-700 009, India.
[3]Department of Physics, Taki Government College, Taki -743429, India.
[4]Department of Physics, Bangabasi Morning College, 19 R. C. Sarani, Kolkata -700 009, India.
[5]Inter-University Accelerator Centre, P.O. Box 10502, Aruna Asaf Ali Marg, New Delhi 110067, India.
[6]Advanced Materials Technology Dept., Institute of Minerals and Minerals Technology, Bhubaneswar – 751013, India.
[7]Department of Physics and Materials Science, Indian Institute of Technology, Kharagpur, India.

We report a systematic study on 1.2 MeV $Ar^{8+}$ irradiated ZnO by x-ray diffraction (XRD), room temperature photoluminescence (PL) and ultraviolet-visible (UV-Vis) absorption measurements. ZnO retains its wurtzite crystal structure up to maximum fluence of $5 \times 10^{16}$ ions/$cm^2$. Even, the width of the XRD peaks changes little with irradiation. The UV-Vis absorption spectra of the samples, unirradiated and irradiated with lowest fluence ($1 \times 10^{15}$ ions/$cm^2$), are nearly same. However, the PL emission is largely quenched for this irradiated sample. Red shift of the absorption edge has been noticed for higher fluence. It has been found that red shift is due to at least two defect centers. The PL emission is recovered for $5 \times 10^{15}$ ions/$cm^2$ fluence. The sample colour is changed to orange and then to dark brown with increasing irradiation fluence. Huge resistivity decrease is observed for the sample irradiated with $5 \times 10^{15}$ ions/$cm^2$ fluence. Results altogether indicate the evolution of stable oxygen vacancies and zinc interstitials as dominant defects for high fluence irradiation.

[*]Corresponding author: djphy@caluniv.ac.in

## 1. Introduction

ZnO is a topic of intense research over the years due to some of its unique properties like wide band gap (~3.4 eV), large exciton binding energy (~ 60 meV), high optical gain and short luminescence life time [1]. Such a material fits for efficient optoelectronic devices [2]. ZnO bears an extreme complex and tuneable defect structure [3]. Properties of ZnO can be tailored by suitable defect engineering. So, generation of defects, their characterization and understanding the defect property interrelations have become an important topic of research [4,5]. In this report, we present a systematic study on 1.2 MeV Ar ion irradiated ZnO. The concomitant changes have been investigated from x-ray diffraction (XRD), room temperature (RT) photoluminescence (PL) and UV-Visible (UV-Vis) absorption measurements. In the low fluence regime, zinc vacancies ($Zn_V$) have been found to be most abundant in the sample. The dominant presence of zinc interstitial ($Zn_I$) and oxygen vacancy ($O_V$) defects can be identified with high irradiation fluence ($> 10^{15}$ ions/cm$^2$).

## 2. Experimental outline

As-supplied ZnO powder (99.99%) have been palletized, annealed in air at 500 °C and cooled slowly (36 °C/h) [6]. Samples have been irradiated by 1.2 MeV $Ar^{+8}$ ion with $1 \times 10^{15}$, $5 \times 10^{15}$ and $5 \times 10^{16}$ ions/cm$^2$ fluence. Microstrain in the samples has been estimated from broadened XRD peaks (recorded in Philips PW 1830 powder diffractometer) constructing Williamson-Hall (W-H) plots [7]. PL measurement has been done with He-Cd laser source (325 nm excitation with power 45 mW). The absorption coefficient, α(λ) has been evaluated from extinction coefficient, k(λ) (measured in Hitachi U-3501 spectrophotometer), using relation α(λ) = 4πk(λ)/λ, λ is wavelength of the absorbed photon [3].

## 3. Results and Discussion

The range of 1.2 MeV Ar ions in ZnO has been calculated (~ 1 μm) using SRIM code [8]. In first 100 nm (from where PL signal comes) electronic energy loss by ionization is very high. Nuclear energy loss dominates in last 200 nm of the trajectory.

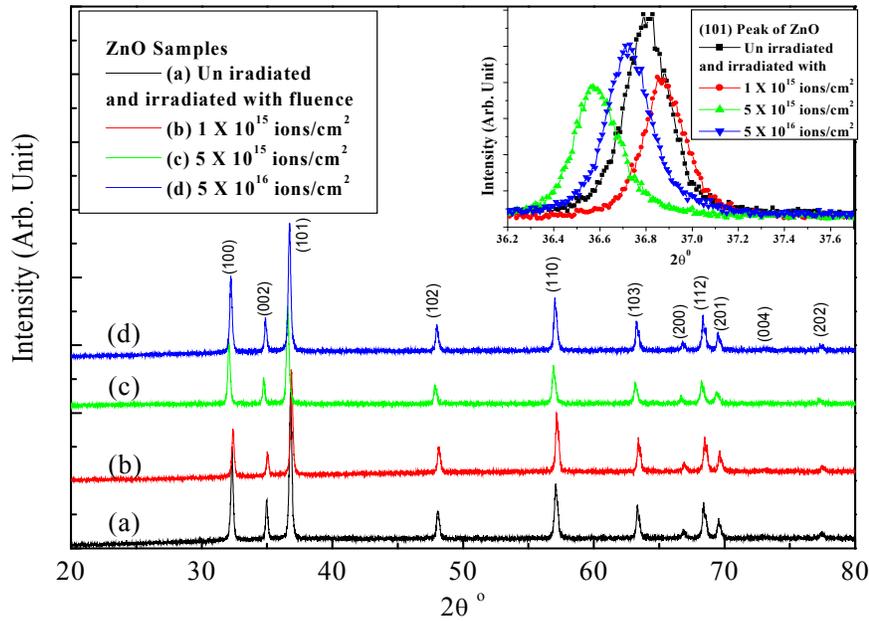

**Figure 1** XRD spectra of the ZnO samples. Inset shows the enlarged view of 101 XRD peak regions.

Possible generation, recovery and re-organization of defects have been discussed in detail in ref. 6. Figure 1 shows the XRD pattern of the samples. It is clear that ZnO remains crystalline even with $5 \times 10^{16}$ ions/cm² fluence. Little change of peak widths and positions with irradiation has been noticed, reflecting possible change of micro-strain in ZnO lattice. Construction of W-H plots is shown in figure 2. The value of strain changes from ~ $2.86 \times 10^{-3}$ (no fluence) to ~ $3.88 \times 10^{-3}$ (highest fluence). The interesting observation is that for fluence $5 \times 10^{15}$ ions/cm$^2$, the W-H plot of the sample is highly scattered and no fitting is possible. This is an evidence of anisotropic strain in the lattice. Possibly, $O_V$s (or complexes) are responsible for such anisotropic strain in ZnO [3]. Homogenization of disorder as well as its saturation is more likely [9,10] in the sample irradiated with highest fluence. In this case, the induced micro-strain almost similar in all crystal directions giving a linear W-H plot.

The evolution of defects with irradiation is more prominent in the absorption

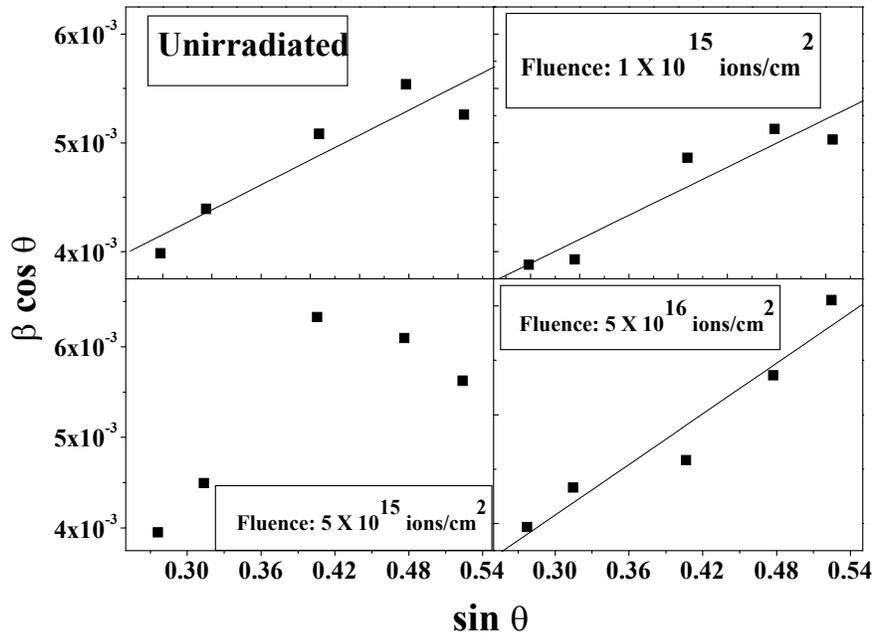

**Figure 2** W-H plots of the ZnO samples.

spectrum of the samples. The maximum absorption is decreased with increasing fluence (Figure 3(a)). Enhanced absorption in defect states and increased microstrain mainly contribute to the reduction of band gap absorption. For irradiation with highest fluence, the value of α is lowest in the whole spectral range (with huge reduction of maximum absorption). To note, the value of the incorporated microstrain is highest for this sample. Moreover, free carrier absorption in the infrared region may also reduce the UV-Vis absorption in this sample. The unirradiated sample and the irradiated one with

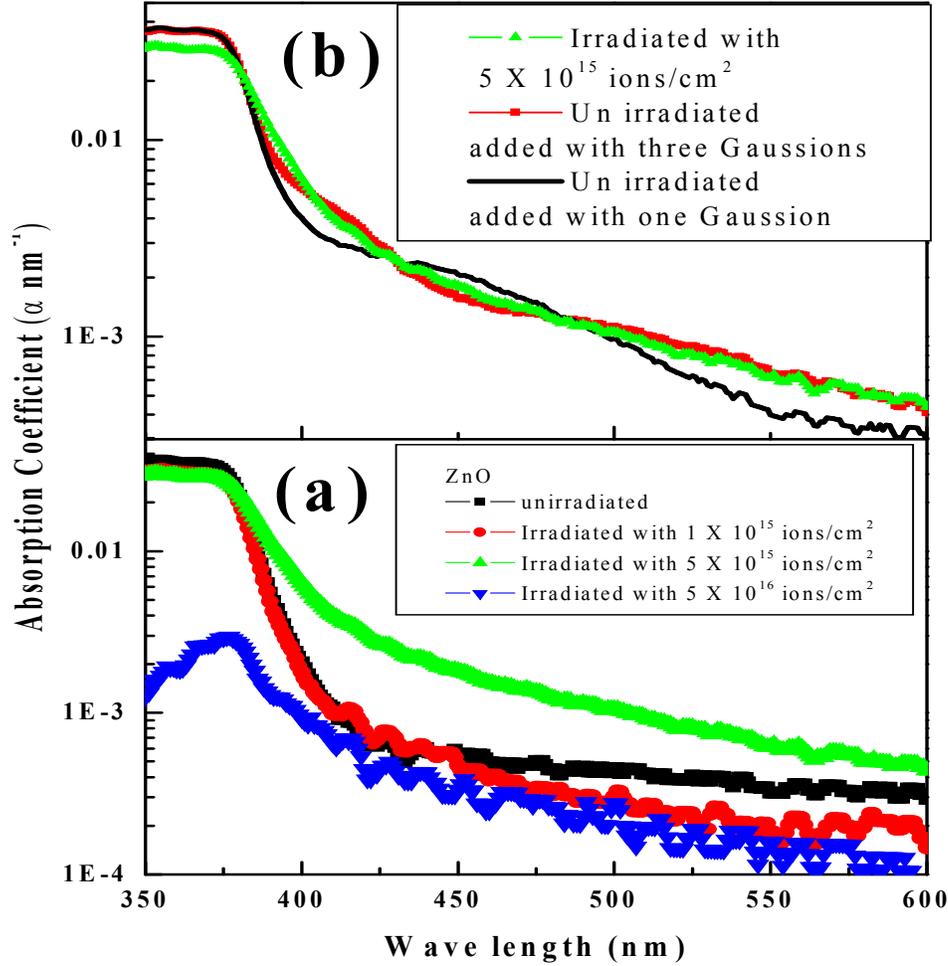

**Figure 3** (a) UV-Vis absorption spectra of the ZnO samples.
(b) Comparison of the experimental and fitted spectra.

lowest fluence show nearly similar absorption edges (~ 3.22eV). The samples irradiated with higher fluence ($5 \times 10^{15}$ and $5 \times 10^{16}$ ions/cm$^2$) show a red shift of this edge (~ 3.19 eV) with pronounced band tailing. Kappers et al. have reported [11] huge shift of absorption edge due to annealing of ZnO in Zn vapor. They have shown that such a shift can be understood as the evolution of a Gaussian shaped absorption band at 3.03 eV (FWHM ~ 0.8 eV) and assigned this band of ZnO with the $O_V$s. We have also observed [3] lowering of band edge and increase of sub-band edge absorption (band tailing) in annealed nano-ZnO. Occurrence of band tailing does not correlate with positron spectroscopic [3] results, which are sensitive to $Zn_V$s. In this study, the spectrum for the un-irradiated sample nearly traces (Fig. 3(b)) that of the irradiated one ($5 \times 10^{15}$ ions/cm$^2$) when we add three Gaussian bands at 3.10 eV (FWHM ~ 0.47 eV), at 2.89 eV (FWHM ~ 0.73 eV) and at 2.48 eV (FWHM ~ 0.76 eV). Adding one Gaussian (at 3.03 eV, like Ref. 11) with unirradiated spectrum, result much worse agreement with experimental spectrum of irradiated sample (Fig. 3(b)). So, it can be inferred that at-least three defect centers have been generated due to high fluence irradiation. Defect states close to the band edges (3.10 and 2.89 eV) are responsible for the reduced absorption edge.

PL measurements show (Fig. 4) quenching of luminescence with initial irradiation. This is, most probably, due to generation of $Zn_V$s, which act as non-radiative defects centers in ZnO [12]. At such fluence (~ $10^{15}$ ions/cm$^2$), isolated $Zn_V$s have been observed in 2 MeV O irradiated ZnO [10]. Separation of generated defects is lower at higher fluence and dynamic recovery of $Zn_V$s becomes more favorable [10]. So, other defects such as $Zn_I$, $O_V$ and antisite defects become dominant. Such defects are luminescent centers in ZnO [4] and luminescence is increased in the whole spectral range [6]. It is also possible that at RT the UV peak (~ 3.27 eV) is an admixture of both free exciton and defect related transitions [6]. In such case, recovery of the UV peak with enhanced disorder may be possible. Finally, the PL yield from ZnO depends on several competitive factors like probability of radiative and non-radiative transitions, nature and abundance of defects, lattice constants and strain etc.. Similar to our results, non-monotonic dependence of PL yield with increasing irradiation fluence has also been observed earlier [13]. However, it is an interesting issue that should be investigated with greater detail.

Besides the UV peak, presence of three other emission peaks ~ 3.17 eV, ~ 2.80 eV and ~ 2.43 eV (close to the absorption bands) can also be identified (Fig. 4). High fluence irradiation causes an increase of the relative weights these peaks compared to the UV peak. PL peaks ~ 2.9 eV and ~ 3.15 eV are generally attributed to $Zn_I$ related defects [14,15]. Also, it has been predicted that energy state of photo-excited $O_V$s in ZnO may lie close to the conduction band minimum [16]. In general, $Zn_I$s are mobile in ZnO [4]. But, in presence of $O_V$, they may become stable as theorized recently [5]. Stable co-existence of $Zn_I$ and $O_V$ defects have also been identified experimentally in annealed ZnO based ceramics [17]. It has been shown [6,18] that luminescence peak ~ 2.4 eV in ZnO is from $O_V$ defects. Here, we find the peaks at 2.80 and 3.17 eV are also $Zn_I$ and $O_V$ related.

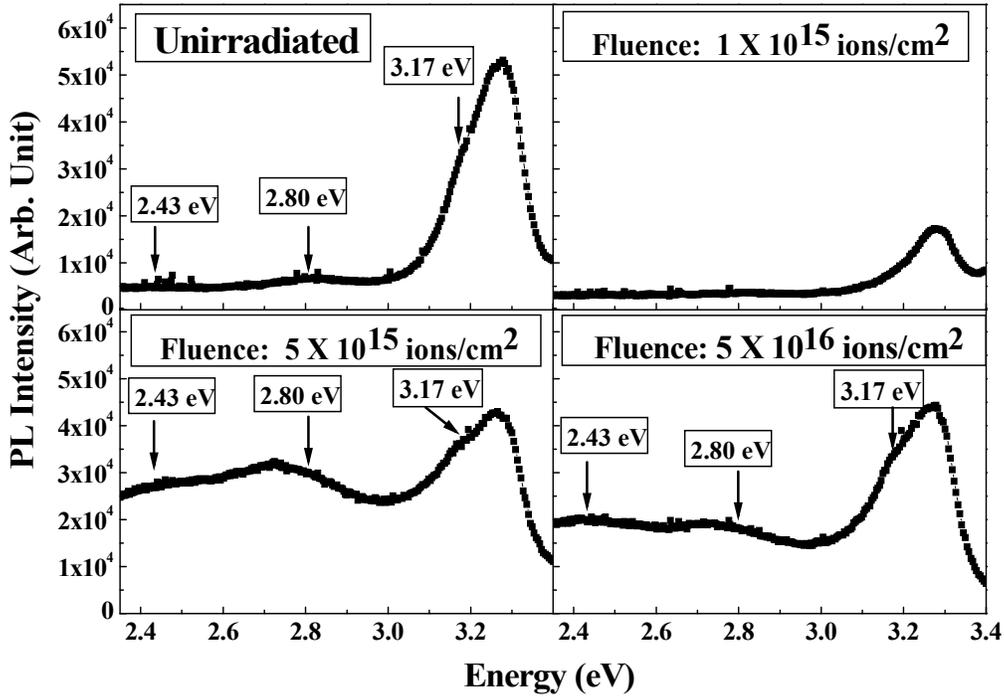

**Figure 4** RT PL spectra of the ZnO samples.

The colour of the samples changes from white to dark reddish brown and resistivity is lowered drastically due to $5 \times 10^{15}$ ions/cm$^2$ fluence [6]. Large reduction of resistivity near the subsurface region has been accounted for Zn-O divacancy clusters with abundant $O_V$s in the N implanted ZnO [18]. Present study confirms the presence of $Zn_I$ and $O_V$s in ZnO irradiated with high fluence of Ar ion. We attribute the simultaneous coloration and lowering of resistivity is due to mutually stabilized $O_V \tilde{} Zn_I$ defect pairs.

## 4. Conclusion

Important understanding regarding the evolution of different defect species in ZnO due to low/medium energy ion irradiation is discussed here. From the red shift of optical absorption edge, generation of at-least two defect centers can be identified. This contention is being supported by XRD and PL results. At high irradiation fluence, most abundant stable defects are $Zn_I$ and $O_V$.

**Acknowledgements**

Financial help from DST-FIST, Govt. of India is acknowledged. Soubhik Chattopadhyay is grateful to Govt. of West Bengal for providing Research Fellowship**.**

**References**

[1] C. Klingshirn, J. Fallert, H. Zhou, J. Sartor, C. Thiele, F. Maier-Flaig, D. Schneider, H. Kalt, Phys. Stat. Sol (b) **247**, 1424 (2010).

[2] S.J. Pearton, D.P. Norton, K. Ip, Y.W. Heo, T. Steiner, Prog. Mater. Sci. **50**, 293 (2005).

[3] S. Dutta, S. Chattopadhyay, A. Sarkar, M. Chakrabarti, D. Sanyal, and D. Jana, Prog. Mat. Sci. **54**, 89 (2009).

[4] A. Janotti and C.G. Van de Walle, Phys Rev B **76**, 165202 (2007)

[5] Y. –S. Kim, and C. H. Park, Phys. Rev. Lett. **102,** 086403 (2009).

[6] S. Chattopadhyay, S. Dutta, D. Jana, S. Chattopadhyay, A. Sarkar, P. Kumar, D. Kanjilal, D. K. Mishra and S. K. Ray, J. Appl. Phys, **107**, 113516 (2010) and references therein.

[7] G. K. Williamson and W. H. Hall, Acta Metall. **1,** 22 (1953)

[8] J.F. Ziegler, J.P. Biersack, U. Littmerk, Stopping Power and Ranges of Ion in Matter, Pergamon Press, New York, 1985.

[9] E. Wendler, O. Bilani, K. Gärtner, W. Wesch, M. Hayes, F. D. Auret, K. Lorenz, and E. Alves, Nucl. Instrum. and Meth. B **267**, 2708 (2009).

[10] A. Zubiaga, F. Tuomisto, V.A. Coleman, H.H. Tan, C. Jagadish, K. Koike, S. Sasa, M. Inoue and M. Yano, Phys. Rev. B **78**, (2008) 035125.

[11] L.A. Kappers, O.R. Gilliam, S.M. Evans, L. E. Halliburton and N. C. Giles, Nucl. Instrum. and Meth. B **266**, 2953 (2008).

[12] T. Koida, A. Uedono, A. Tsukazaki, T. Sota, M. Kawasaki, and S.F. Chichibu, Phys. Stat. Sol. (a) **201**, 2841 (2004).

[13] H. Zang, Z.G. Wang, X.P. Peng, Y. Song, C.B. Liu, K.F. Wei, C.H. Zhang, C.F. Cao, Y.Z. Ma, L.H. Zhou, Y.B. Sheng and J. Gou, Nucl. Instrum. and Meth B **266**, 2863 (2008).

[14] N.S. Han, H.S. Shim, J.H. Seo, S.Y. Kim, S.M. Park and J.K. Song, J. Appl. Phys. **107**, 084306 (2010).

[15] B. Cao, W. Cai, and H. Zeng, Appl. Phys. Lett. **88**, 161101 (2006).

[16] S. Lany and A. Zunger, Phys. Rev. Lett. **98**, 045501 (2007).

[17] P. Cheng, S. Li, L. Zhang and J. Li, Appl. Phys. Lett. **93**, 012902 (2008).

[18] Y. Dong, F. Tuomisto, B.G. Svensson, A. Yu Kuznetsov and L.J. Brillson, Phys. Rev. B **81**, 081201 (R) (2010).